\documentclass{IEEEtran}
\usepackage{cite}
\usepackage{amsmath,amssymb,amsfonts}
\usepackage{graphicx}
\usepackage{textcomp,nicefrac}
\usepackage{url}
\usepackage{lineno}
\def\BibTeX{{\rm B\kern-.05em{\sc i\kern-.025em b}\kern-.08em
T\kern-.1667em\lower.7ex\hbox{E}\kern-.125emX}}
\begin{document}
\title{Curation and Dissemination of Complex Multi-modal Data Sets for Radiation Detection, Localization, and Tracking}
\author{Nicolas Abgrall, Mark S. Bandstra, Reynold J. Cooper, Marco Salathe, Brian J. Quiter, Rajesh Sankaran, Yongho Kim, and Sean Shahkarami
\thanks{This paper was submitted for review on August 8, 2025. This work was performed under the auspices of the U.S. Department of Energy by Lawrence Berkeley National Laboratory (LBNL) under Contract DE-AC02-05CH11231. The project was funded by the U.S. Department of Energy, National Nuclear Security Administration, Office of Defense Nuclear Nonproliferation Research and Development.
N. Abgrall is with the Lawrence Berkeley National Laboratory, Berkeley, CA 94720 USA (e-mail: nabgrall@lbl.gov).}
\thanks{M.S. Bandstra (e-mail: msbandstra@lbl.gov), R.J. Cooper (e-mail: rjcooper@lbl.gov), M. Salathe (e-mail: msalathe@lbl.gov), B.J. Quiter (e-mail: bjquiter@lbl.gov) are with the Lawrence Berkeley National Laboratory, Berkeley, CA 94720 USA.}
\thanks{R. Sankaran (e-mail: rajesh@anl.gov), Y. Kim (e-mail: yongho.kim@anl.gov), and S. Shahkarami (e-mail: sshahkarami@anl.gov) are with the Argonne National Laboratory, Lemont, IL 60439 USA.}
}
\maketitle

\begin{abstract}
The PANDAWN sensor network in Chicago, IL, is a state-of-the-art test-bed for networked, multi-modal sensing. It integrates AI/data science methods into its operation, from data acquisition to automated data labeling and curation workflows. The curation and dissemination of diverse multi-modal data sets will enable the development of new radiological/nuclear (R/N) detection, localization, and tracking algorithms, and methods relevant across the nonproliferation mission space. This paper first introduces the PANDAWN sensor network and the features that make it stand out from previous multi-modal data acquisition efforts. We then review the various data streams acquired on the PANDAWN nodes, and present the implementation of an automated data curation pipeline that includes the labeling of radiation and contextual data streams. We finally provide a short overview of different studies that leveraged the curated data sets.
\end{abstract}

\section{Introduction}
\label{sec:introduction}
\begin{figure}[t]
\centerline{\includegraphics[width=3.6in]{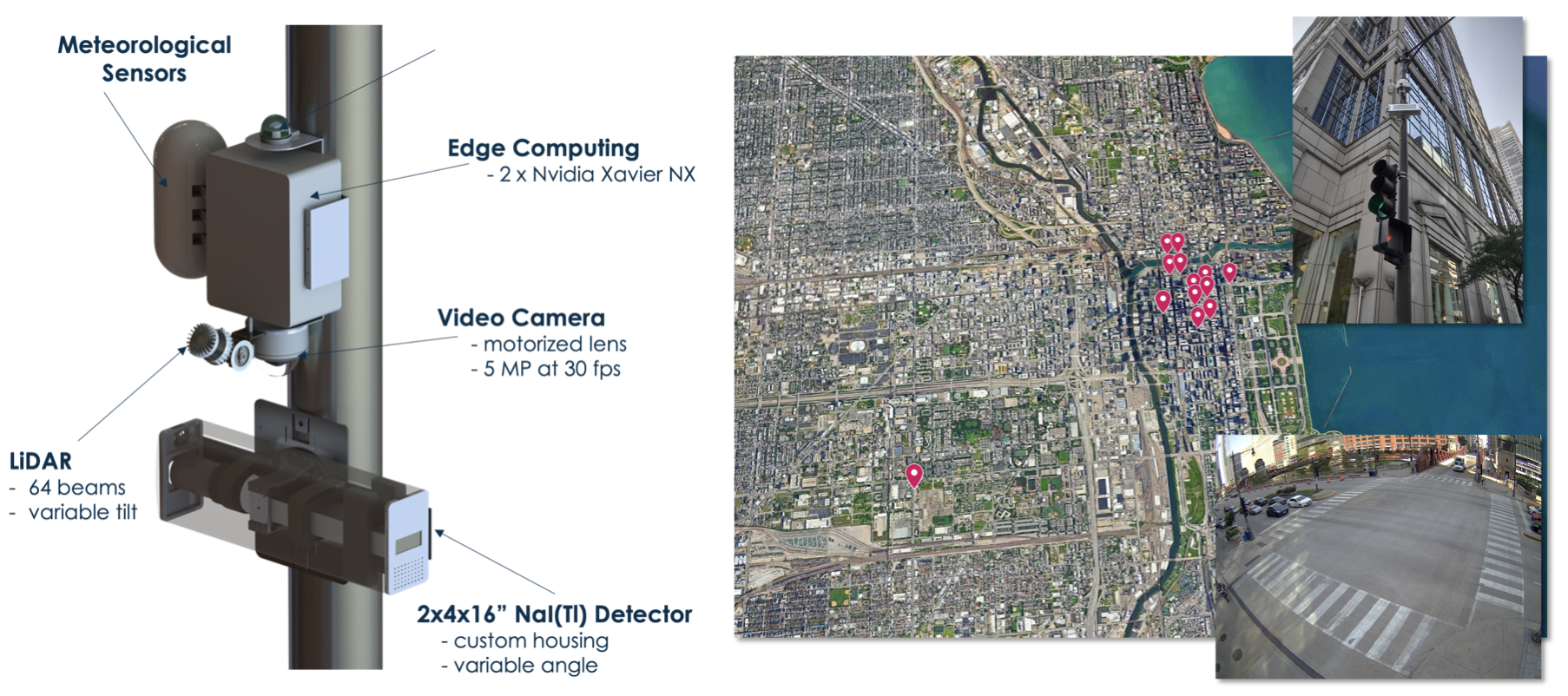}}
\caption{Engineering drawing of a PANDAWN node (left), and map showing the deployment locations of 13 such nodes in Chicago, IL (right). Inserts show a node mounted on a traffic light pole (top), and a typical intersection field of view (bottom).}
\label{PANDAWN_nodes}
\end{figure}

The PANDAWN sensor network (\figurename~\ref{PANDAWN_nodes}) is a testbed comprising 13 statically deployed and networked multi-modal radiation detection systems (or nodes) in Chicago, IL. The PANDAWN nodes were developed and fielded in coordination between the Lawrence Berkeley National Laboratory (LBNL)-led Platform and Algorithms for Networked Detection and Analysis (PANDA) project, and the Argonne National Laboratory (ANL)-led Domain Aware Waggle Network (DAWN) project~\cite{RC}. The sensing modalities supported by each node include a 2x4x16" NaI(Tl) gamma-ray detector, a 64-beam LIDAR and a 5 megapixel camera for contextual awareness, and a suite of meteorological sensors (temperature, pressure, relative humidity) and a rain gauge for environmental information. The sensing modalities are complemented by 2 NVIDIA Jetson Xavier NX boards for edge computing capabilities, while networking provides additional cloud processing capabilities. Edge computing is a distributed paradigm that processes data near the devices or instruments where it is generated (in-situ), rather than relying solely on centralized cloud servers, thereby reducing latency, improving responsiveness. Because data can be processed and discarded locally at the edge without being stored or transmitted, the approach also supports privacy-preserving sensing, which is especially important in urban deployments where privacy is a significant concern. The PANDA project provided the algorithmic development architecture and radiological and contextual data processing methods, and the DAWN project provided the hardware platform, and programmable edge computing, and networking infrastructures. The deployment of the first 13 nodes of the network was completed in August 2023, and a first acquisition campaign took place from September 2023 to May 2024. A second acquisition campaign is expected to start by the end of 2025, possibly including additional nodes. 

Other detector arrays have been previously deployed in urban environments to support nonproliferation missions. Original efforts, though, have concentrated on radiation detection only. In particular, NoVArray was an array of nineteen 2x4x16" NaI(Tl) detectors deployed on traffic light poles at major traffic intersections throughout Northern Virginia~\cite{NH}. The detectors collected 1024-bin spectra up to 3 MeV at a rate of 1 Hz, and reported these spectra along with time, and GPS coordinates. The SIGMA project from DARPA~\cite{SIGMA}, is another example of deployment, with similar gamma-ray detectors deployed on the fleet of Washington D.C. Fire and Emergency Medical Services ambulances for approximately seven months. The Mobile Imaging and Spectroscopic Threat Identification (MISTI) system~\cite{MISTI}, whose primary goal was to measure background variability, performed radiation background measurements across a portion of the continental United States with an array of twenty-eight high purity germanium detectors. While these projects provided a rich gamma data set for the nonproliferation and data science communities, they lacked contextual information which enables the development of more advanced algorithms and provides a level of ground truth.

Several multi-modal sensing development efforts provided richer, more complex data sets enabling radiation and contextual data fusion for the development of detection algorithms in urban environments. While the primary focus of the Multi-agency Urban Search Experiment (MUSE)~\cite{MUSE-1}\cite{MUSE-2} was on accurately modeling urban radiological environment for synthetic data generation, several dedicated multi-modal measurement campaigns were performed to support thorough benchmarking of the simulation developments. In particular, a measurement campaign developed a full radiological characterization of a controlled facility that roughly corresponded to two city blocks. As part of the campaign, the Radiological Multi-sensor Analysis Platform (RadMAP)~\cite{RadMAP-1} collected extensive multi-sensor data~\cite{RadMAP-2}. The RadMAP system was developed as a testbed to explore correlations between radiological and contextual data, by combining large volume gamma-ray and neutron detectors with contextual sensors such as panoramic video, LIDAR, and hyperspectral imagery. 
As part of another campaign, a multi-year dataset was acquired from a stationary detector system named MUSE01 located at the entrance to the High Flux Isotope Reactor / Radiochemical Engineering Development Center (HFIR/REDC) at Oak Ridge National Laboratory (ORNL)~\cite{MUSE-3}. The system consists of a variety of sensors including a 2x4x16" NaI(Tl) detector, a weather station, a LIDAR, and HVAC (heating, ventilation, and air conditioning) for temperature stability. In particular, this dataset was used in an effort to analyze and model gamma-ray spectra measured during rainfall in order to derive accurate source terms for $^{214}$Pb and $^{214}$Bi~\cite{MB-2}, and thus provide better account of precipitation induced effects to mitigate false alarms in radiological contamination/source search scenarios. While all those data sets allow to explore and leverage correlations between radiation data and contextual information, they are limited to single systems deployed in test environments. 

The Multi-Informatics for Nuclear Operations Scenarios (MINOS) project~\cite{MINOS} was established to develop methods that leverage persistent monitoring data collected in and around ORNL's HFIR/REDC to make determinations about the ongoings at a nuclear facility. The general approach of MINOS was to fuse data collected from multiple sensing modalities in order to
characterize operations occurring at nuclear facilities. Sensing devices measured electromagnetic, radiation, thermal, seismic, infrasound and low-frequency acoustics from nuclear facility operations. Although it is not focused on urban environments, MINOS does combine multi-modal sensing with a networked array configuration, with multi-modal sensing platforms
deployed around the facility.

All these projects provided useful data sets, however, the PANDAWN network combines a unique set of features enabling a more complete exploration of radiation detection and localization in urban environments. Indeed, the PANDAWN network leverages well understood and characterized sensors operating within a software environment that provides robust synchronization of all data streams. The PANDAWN nodes have been deployed at fixed location in a real urban environment, for long-dwell time data collection. Additionally, the ability to process data at the edge, networking, and cloud-based processing allow for informed, adaptive data acquisition of specific data sets, at both single detector and network levels. Finally, PANDAWN supports automated labeling of all its data streams as part of data curation for the nonproliferation and data science communities. 

\section{Data streams}
\label{sec:data_streams}
\begin{figure}[t]
\centerline{\includegraphics[width=3.5in]{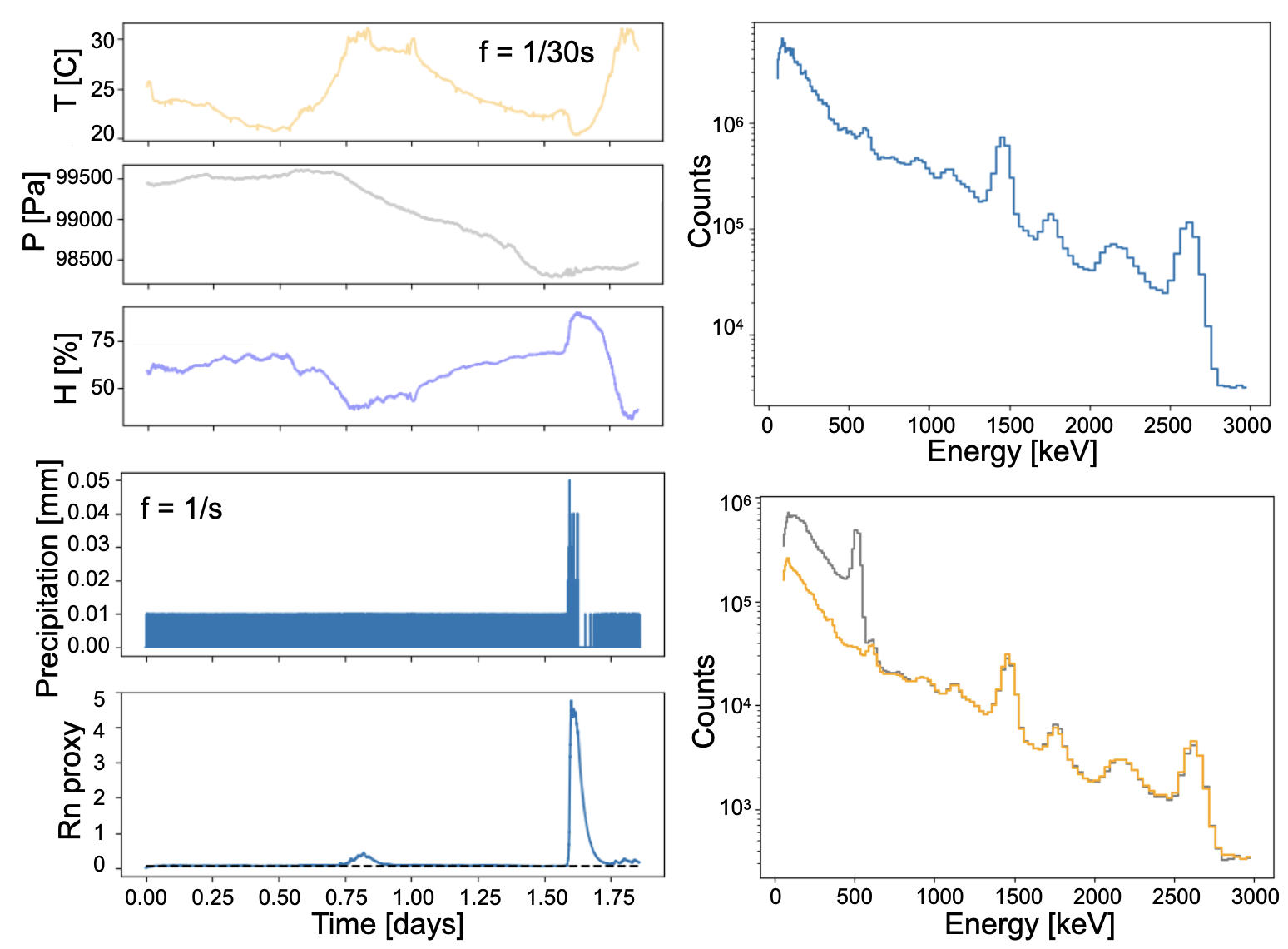}}
\caption{Example time series of environmental data streams for meteorological data (top left), and precipitation and radon progeny proxy (bottom left). Typical static background data (top right), and example of radiation data including the presence of a positron-emitting radiotracer, \textit{e.g.}, $^{18}$F (bottom right).}
\label{rad_env_data}
\end{figure}

The environmental data streams are sampled every 30 seconds for the temperature, pressure, and relative humidity readings, and every second for the precipitation readings from the rain gauge (\figurename~\ref{rad_env_data}). 
The higher sampling frequency of the rain gauge data allows for accurate modeling of the time evolution of the deposition of rain-induced $^{222}$Rn progeny including $^{214}$Pb and $^{214}$Bi. The latter is used as a proxy during background model learning to flag radiation data where spectral features from the radon progeny are expected to be enhanced~\cite{MB}. The radiation data from the  NaI(Tl) detector are acquired as list-mode data, and provided as raw ADC values, as well as calibrated energy values obtained from an automated continuous calibration procedure running at the edge (detailed in Sec.~\ref{sec:curation}). 
While most radiation data samples will predominantly reflect the local background radiation, the thirteen nodes deployed at different intersections significantly increase the encounter probability of typical nuisance (i.e., non-threat) sources expected in urban environments. These include medical imaging isotopes (e.g., $^{99m}$Tc and $^{18}$F) and industrial isotopes (e.g., $^{137}$Cs used in density gauges). An example of an encounter with $^{18}$F is shown in ~\figurename~\ref{rad_env_data}. Thus these data sets are well suited for the development and training of R/N detection and identification algorithms. The environmental and radiation data streams represent about 2~GB/node/day of data transmitted to ANL's datastores.

The contextual data streams from the camera and LIDAR are significantly larger in size and cannot be streamed continuously due to two constraints: (1) bandwidth limitations, and (2) privacy protection policies that restrict the duration of continuous video capture from the camera. To address these challenges, tailored acquisition schemes were implemented to collect datasets suitable for algorithm development and training. For the camera stream, 2-minute long periods of footage are acquired at a rate of 30 frames per second 6 times per day per node, from 2:30~am to 10:30~pm. Such data sets provide a large diversity of environmental and lighting conditions appropriate for the development and training of object detection, identification, and tracking algorithms. 

\begin{figure}[t]
\centerline{\includegraphics[width=3.5in]{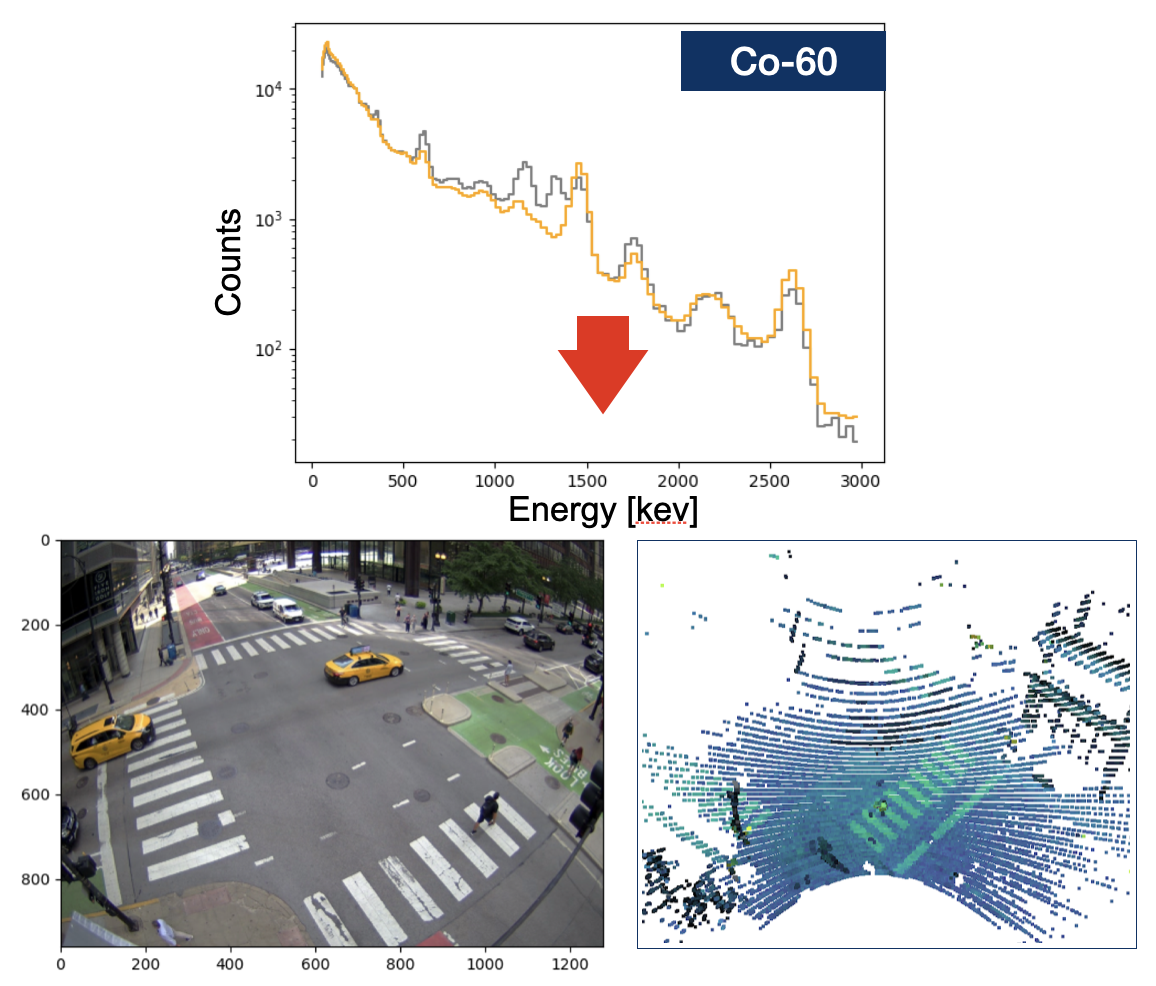}}
\caption{Illustration of the triggered acquisition scheme: upon an R/N anomaly detection (top, \textit{e.g.}, $^{60}$Co detection), high rate camera and LIDAR data (bottom left and right resp.) are synchronously acquired with radiation and environmental data.}
\label{triggered_acquisition}
\end{figure}

Triggered acquisition is another scheme that was developed later on but was not ready for deployment during the time of the first data collect. In this approach, high-rate camera and LIDAR data are synchronously collected alongside radiation and environmental data upon the detection of radiological or nuclear (R/N) features of interest. Data acquisition continues as long as the R/N feature persists, enabling targeted collection of contextual information during relevant events. Triggered acquisition is illustrated in \figurename~\ref{triggered_acquisition} showing typical camera footage and LIDAR point cloud data. This acquisition scheme enables the generation of correlated datasets that are uniquely suited for the development and training of anomaly detection, localization, and tracking algorithms. The contextual data streams currently account for approximately 6~GB/node/day of selected camera footage transmitted to ANL's databases. The implementation of triggered acquisition is expected to increase the data volume by an additional 2~GB/node/day.

\section{Automated data labeling and curation pipeline}
\label{sec:curation}
\begin{figure}[h]
\centerline{\includegraphics[width=3.7in]{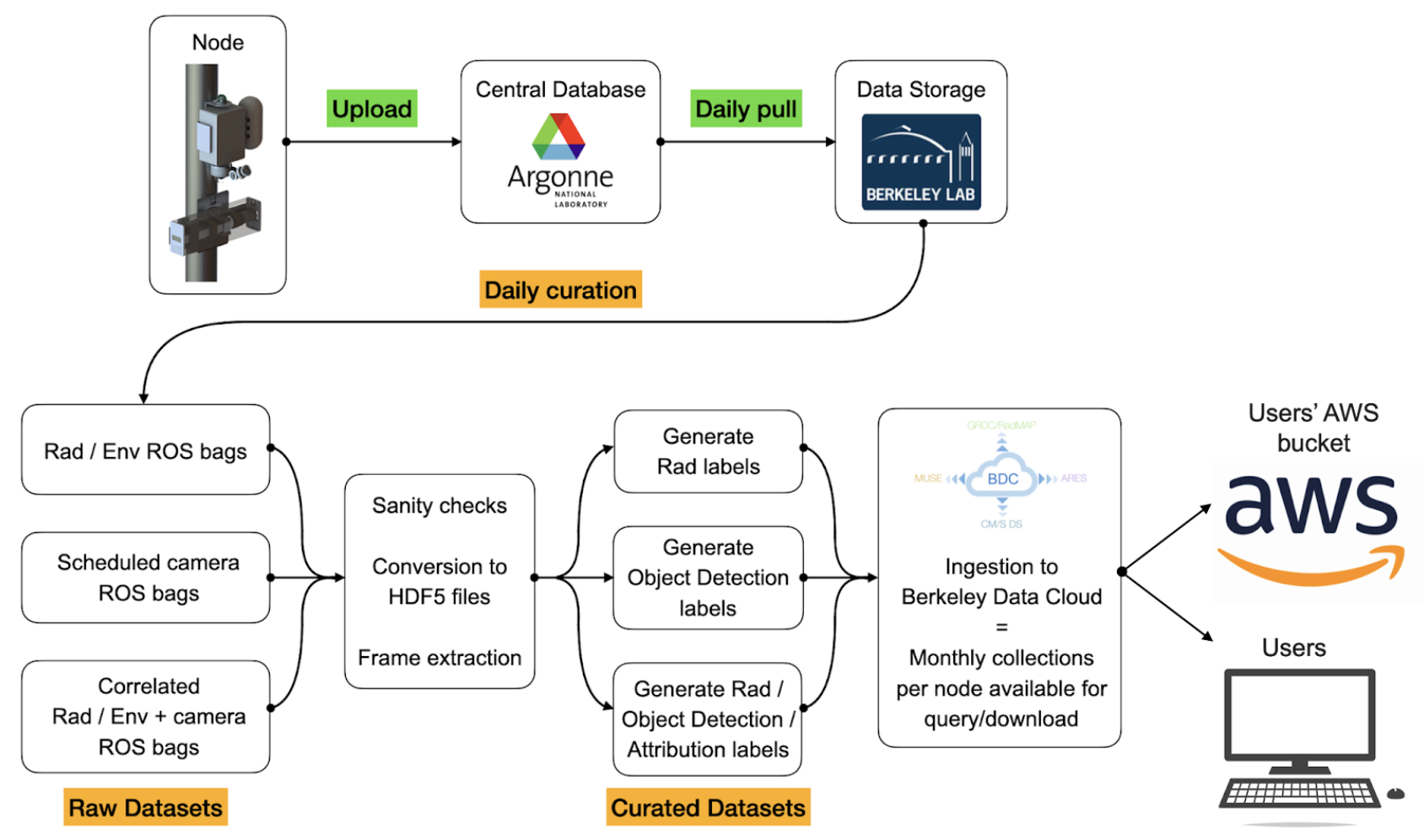}}
\caption{Schematic diagram of the automated labeling and curation workflow.}
\label{curation_pipeline}
\end{figure}

To facilitate the development of advanced radiation detection and data fusion algorithms, an automated pipeline (\figurename~\ref{curation_pipeline}) was implemented to label and curate the data from the PANDAWN sensor network. 

\noindent The pipeline operates on a daily basis and performs the following tasks:
\begin{enumerate}
    \item Pull data from ANL's database to storage at LBNL
    \item Run QA/QC checks on raw data
    \item Run conversion from the Robot Operating System (ROS)~\cite{ROS} format to the Hierarchical Data Format (HDF)~\cite{HDF5}, commonly used in academic research
    \item Generate labels for the radiation, environmental, and contextual data streams
    \item Index and ingest data and labels in monthly collections on the LBNL's Berkeley Data Cloud (BDC)~\cite{BDC-1}\cite{BDC-2} 
\end{enumerate}

The created monthly collections per node are available to users for direct download or possible upload to Amazon Web Services (AWS)~\cite{AWS} buckets for cloud processing. The curated data set, as of July 2025, contains 9 months of data from 13 nodes, including gamma-ray data in list/bin-mode formats, meteorological data (temperature, pressure, relative humidity, precipitation), selected camera footage, and limited R/N and contextual labels (ongoing processing).

\begin{figure}[h]
\centerline{\includegraphics[width=3.in]{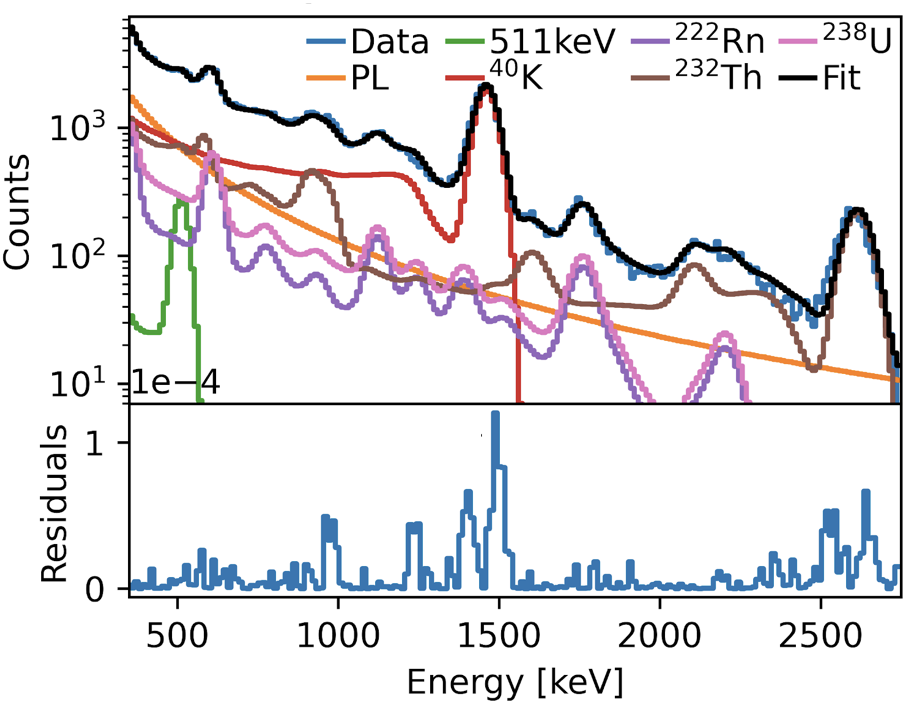}}
\caption{Example calibration fit combining a detector/readout model and a series of simulated NORM templates.}
\label{calibration_fit}
\end{figure}

\begin{figure}[t]
\centerline{\includegraphics[width=3.5in]{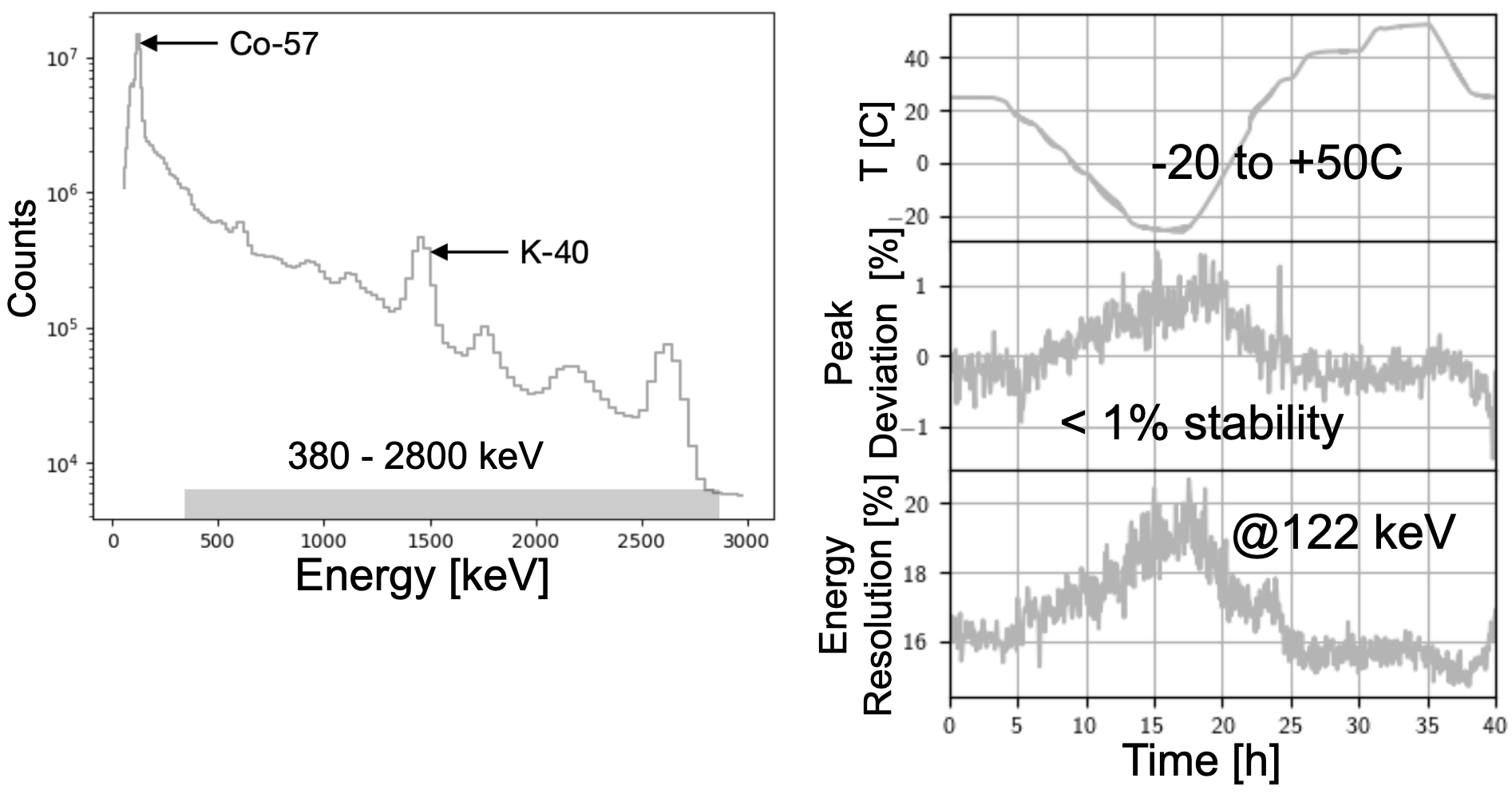}}
\caption{Spectrum acquired during controlled environment testing showing the background 1460~keV peak from $^{40}$K within the fitting range (shown as shaded region), and the 122~keV peak from a $^{57}$Co source below the fitting range (left). A peak stability better than 1\% is observed over the full energy range, shown here at 122~keV, over a temperature sweep from -20 to +50C (right).}
\label{calibration_test}
\end{figure}

The automated labeling step of the curation workflow leverages several algorithms originally developed for online real-time processing at the edge. In particular, the radiation data labeling relies on the automated continuous calibration procedure that runs on the PANDAWN nodes, as well as on the non-negative matrix factorization (NMF) algorithm applied to background learning and anomaly detection~\cite{KB}\cite{MB}. The calibration consists of a combined fit of a detector/readout model, and a series of simulated spectral templates for normally occurring radioactive materials (NORMs). The detector model folds in the crystal light-yield parametrization and energy resolution with a parametrization of the readout electronics (gain, saturation, offset). The NORM templates include $^{40}$K, the uranium and thorium series, radon ($^222$Rn daughters), the 511 keV line from positron-electron annihilation, and an overall power law spectrum capturing the effect of cosmic rays, as shown in \figurename~\ref{calibration_fit}.
By combining a detailed physical model of the detector response with a full-spectrum analysis using background radiation templates, the calibration procedure eliminates the need for an active temperature/gain stabilization. The calibration procedure shows excellent performance (\figurename~\ref{calibration_test}) with peak stability better than 1\%, under various environmental conditions tested in controlled environment, and over the full energy range, in particular, below the fitting range at low energy where non-linearity effects commonly affect calibration.

\begin{figure}[h]
\centerline{\includegraphics[width=3.5in]{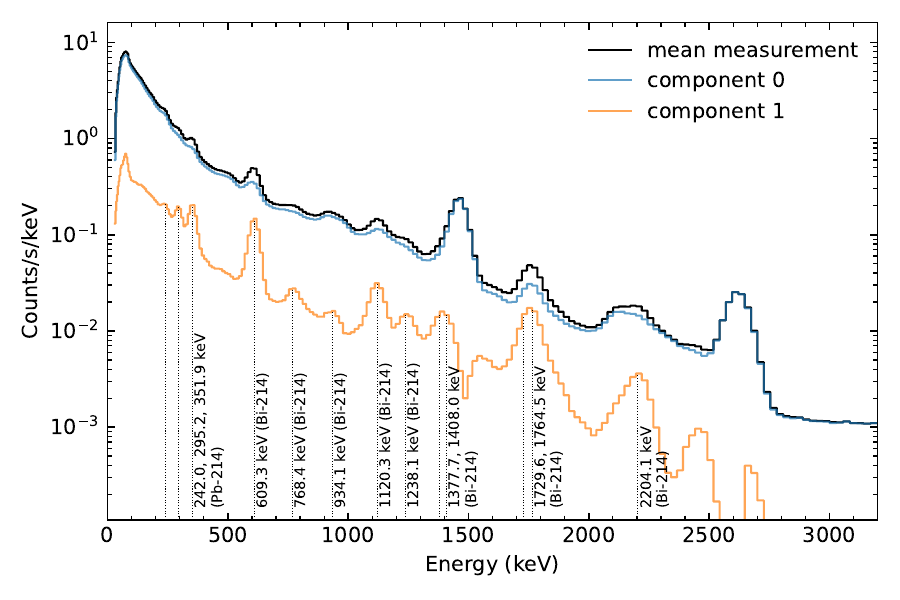}}
\caption{Mean background spectrum, $X$, of dimension (1, d), and basis of NMF components, $V$, of dimension (2, d), showing a component associated to the mean static background, and another associated to the rain-induced radon-enhanced background. The original data, $X$, can be reconstructed as a linear combination, $\hat{X}$, of the NMF components with weights, A, of dimension (1, 2), i.e., $X\approx\hat{X} = AV$, with $A \in \mathbb{R}^{n \times k}_{\geq 0}$, $V\in\mathbb{R}^{k \times d}_{\geq 0}$. Figure from~\cite{MB}.}
\label{nmf}
\end{figure}

\noindent NMF is a data reduction technique similar to principal component analysis (PCA), with the difference that the components are required merely to be positive, but not necessarily orthogonal.~\cite{DL}. As such, NMF lends itself well to the description of radiation data, and is used to learn statistical representations of radiation background components that naturally correlate to some physics interpretations (\figurename~\ref{nmf}). These components are further used for R/N anomaly detection and isotope identification~\cite{KB}.

The labeling of the radiation and environmental data streams is performed using the Background Adaptive Radiation Detection (BARD) package~\cite{NA-1}, which integrates the calibration and NMF algorithms with a spectral triage algorithm~\cite{MB} into an automated background model learning procedure, complemented with automated background model selection, NMF-based R/N anomaly detection, isotope identification, and source encounter generation. BARD generates labels on bin-mode data using a fixed 0.5~s integration time. Labels are created or interpolated at the first timestamp of the integration range, and include:
\begin{itemize}
    \item Timestamp, and bin-mode data,
    \item Live time, and count rate,
    \item Temperature, pressure, and relative humidity, 
    \item Precipitation, and radon proxy,
    \item Spectral triage flag (static, radon, or anomaly) and associated metrics,
    \item NMF weights of the currently selected background model,
    \item Anomaly detection flag (alarm or not) and associated metrics,
    \item Isotope identification flag (alarm or not), isotope IDs, and associated metrics,
    \item Currently selected background model
\end{itemize}

\begin{figure}[h]
\centerline{\includegraphics[width=3.5in]{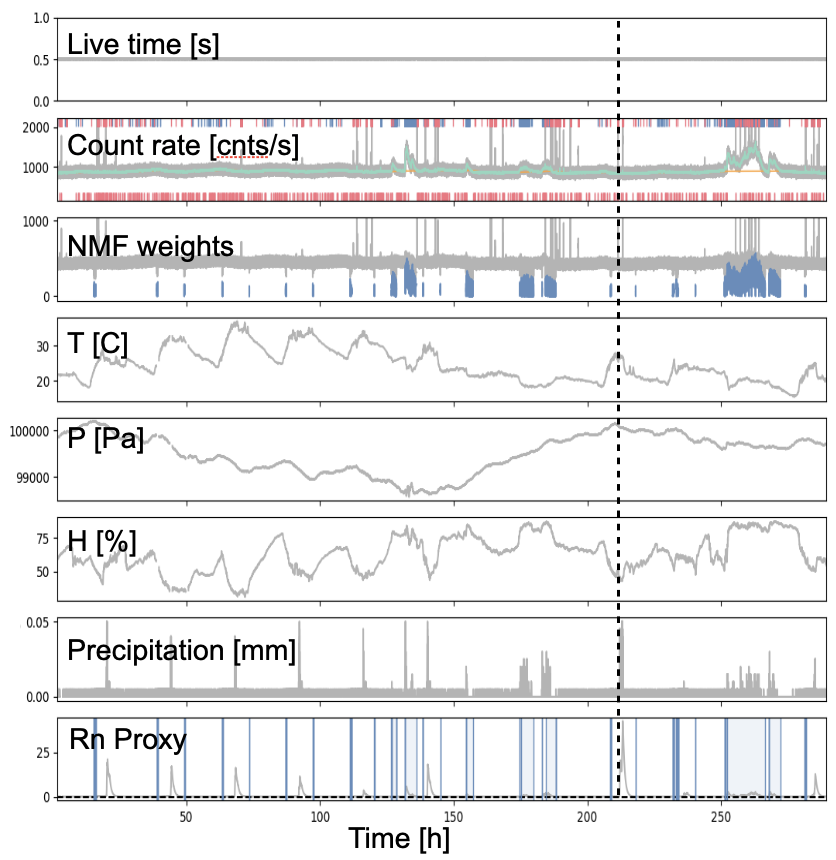}}
\caption{Example time series of radiation and environmental labels automatically generated by the BARD package. Each vertical slice corresponds to the labels generated for a 0.5~s integrated spectrum.}
\label{labels_timeseries}
\end{figure}

\begin{figure}[h]
\centerline{\includegraphics[width=3.5in]{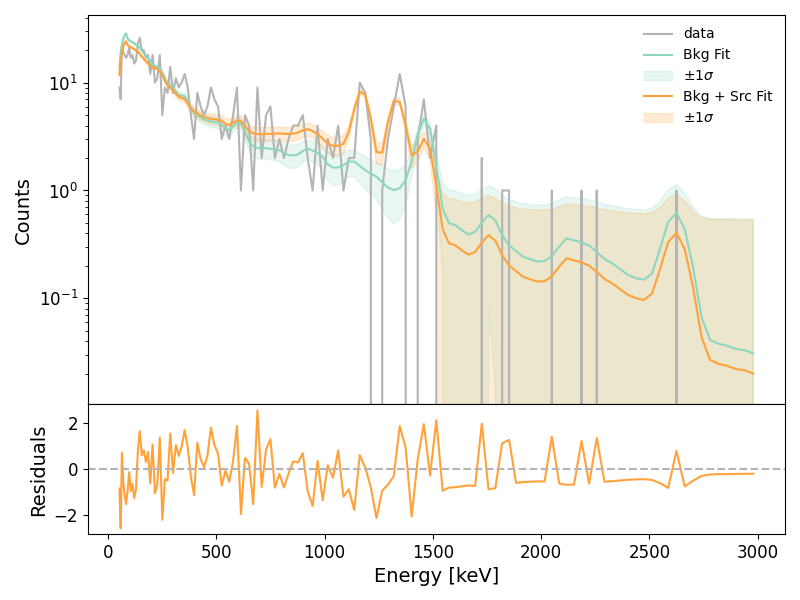}}
\caption{Labels for isotope identification include best guess at isotope ID, and the metrics and background model necessary to visualize the fit of the background-only and background-plus-source hypotheses. Example shows labels associated to a $^{60}$Co identification.}
\label{labels_anomaly}
\end{figure}

Example labels are shown in \figurename~\ref{labels_timeseries} and \figurename~\ref{labels_anomaly} as time series and spectral information respectively. Source encounters are generated upon a first anomaly detection, assigned a unique ID, and last as long as successive anomaly detections are within 10~s of each other, beyond which another encounter ID is generated. Encounter metadata include the encounter ID, duration, indices of the first and last spectra of the encounter, number of alarms during the encounter, as well as our best guess for the encounter isotope ID. The latter is defined as the isotope ID with the largest metric above threshold (likelihood-ratio test between background-only and background-plus-source hypotheses) over the whole set of alarms included in the encounter. All relevant information (NMF components and weights for the extended background-plus-source model, covariance matrices, goodness-of-fit metrics) are included as labels as well for explainability.   
The labeling of the radiation and environmental data streams generates a daily label (HDF5) file including all bin-mode labels, encounter metadata, as well as the background model(s) used for R/N anomaly detection and isotope identification.

\begin{figure}[h]
\centerline{\includegraphics[width=3.in]{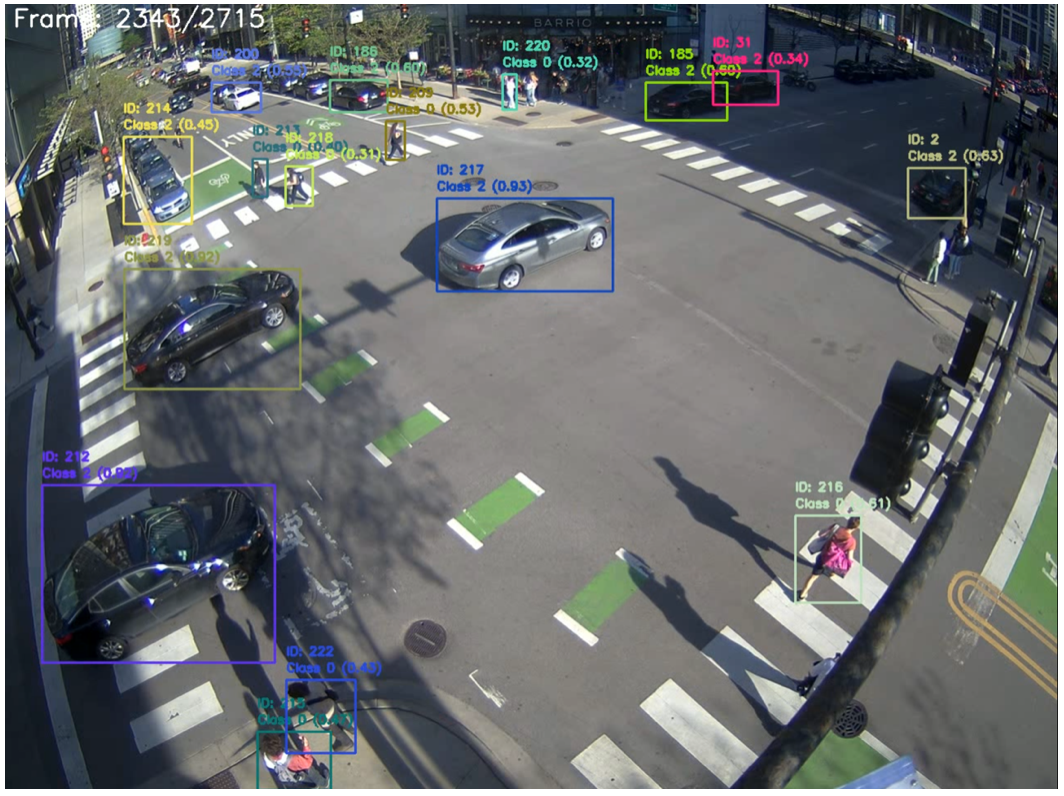}}
\caption{Example of contextual labels superimposed on camera frame, showing a bounding box with class ID, confidence score, and track ID for objects detected in the scene.}
\label{cam_labels}
\end{figure}

The labeling of the contextual data streams is currently limited to selected camera footage, and consists in generating labels for object detection and identification on a per-frame basis, as well as object tracking labels on a frame-to-frame basis. While manual labeling would result in the highest ground-truth accuracy, it is impractical to implement at large scale. Thus, object detection and identification is performed using the YOLOv10~\cite{YOLO} model, with identified classes limited to pedestrians, bicycles, motorcycles, cars, and trucks. Object tracking in image space is performed using the Norfair tracker from Tryolabs~\cite{TRYO}, which assigns a unique identifier to detected objects over the duration of the footage. The inherent uncertainty associated with model-generated labels needs to be kept as low as possible to avoid propagation to other models trained on such labels. To minimize ground-truth errors, the higher accuracy, large model version of YOLOv10 is used. The latter requires larger compute resources compared to smaller versions of the model, which is not a problem for offline processing. In addition, a minimum confidence score of 25\% is required on class identification. While the generated labels are provided as ground-truth, additional selection can still be made to limit error propagation, e.g., higher confidence threshold, discard non-tracked objects. The contextual labels thus include:
\begin{itemize}
    \item Bounding boxes for detected objects,
    \item Class ID and confidence score,
    \item Track ID
\end{itemize}
Additional track metadata are generated and include live time, class ID, and information necessary to reconstruct tracks in image space over the duration of the footage. An example of contextual labels is show in \figurename~\ref{cam_labels}.

As mentioned in Section~\ref{sec:data_streams}, the triggered acquisition scheme allows to capture correlated data from the radiation and camera/LIDAR data streams upon R/N anomaly detection. While not deployed during the first acquisition campaign, tools were developed to perform source-object attribution~\cite{MM} to label such data sets as well. The attribution algorithm executes the following steps:
\begin{enumerate}
    \item Reconstruction of the 3-D trajectories of detected objects from image to detector space
    \item Modeling of expected energy dependent signal, $c_i(E)$, at each point, $i$, of the trajectories assuming sources emitting an isotropic flux, $\alpha$, are carried along:
        \begin{equation}
        \label{eq1}
            c_i(E)=\frac{\epsilon(\Omega, E)\alpha e^{-\mu(E) r_i}}{4\pi {r_i}^2}\Delta t_i
        \end{equation}
    where $\epsilon$ is a solid angle and energy dependent detector efficiency term, $r$ is the euclidean distance from the detector center to the tracked object, $\mu$ a material (air) and energy dependent attenuation coefficient, and $\Delta t_i$ the dwell time at position $i$.
    \item Fit of the signal model to the corresponding time series of the NMF weights of the identified isotope
    \item Attribution of signal to most probable object based on maximum figure-of-merit (p-value associated to $\chi^2$ of the fit)
\end{enumerate}

\begin{figure}[h]
\centerline{\includegraphics[width=3.in]{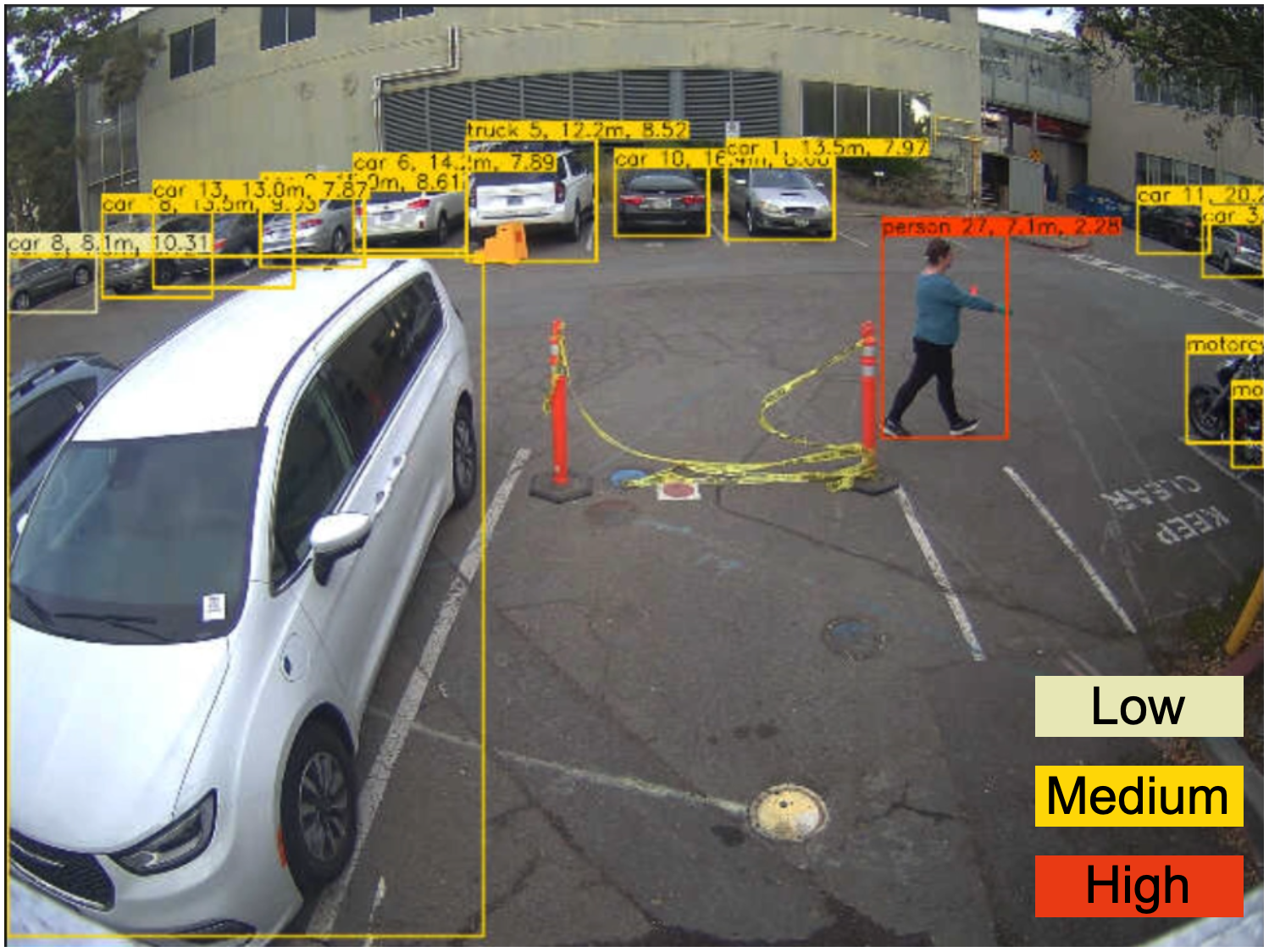}}
\caption{Example source-object attribution labels generated for a pass-by with a 200$~\mu$Ci $^{60}$Co source, showing the likelihood of attribution of tracked objects in the scene. The trajectory followed by the source carrier (orange bounding box) results in the highest attribution score.}
\label{soa_labels}
\end{figure}

\section{Example studies}
\label{sec:examples}
To illustrate how the labeled and curated data sets presented in this paper can be used, we mention in this section a couple studies that are either directly based on these data sets, or have used them to validate simulations, both at the node and network levels:
\begin{itemize}
    \item Spectral shape clustering~\cite{MB}: explores various clustering methods to learn isotope-specific spectral shape templates to be used in place of simulated templates for isotope identification
    \item Adaptive NMF~\cite{CJ}: implements a time dependent feedback loop over an NMF-based background learning procedure, including automatic selection and refitting of the models over time
    \item Network-wide optimization of context-based rad data fusion~\cite{ER}\cite{NA-2}: optimizes detector placement and the fusion of the radiation data streams from different sensors of the network to increase the signal-to-noise ratio and anomaly detection probability, as a function of the level of contextual information that can be shared over the network
    \item Characterization of rain-induced radon progeny depositions at city-scale~\cite{SD}: studies the city-scale topology and time evolution of radon progeny depositions induced by rain events 
\end{itemize}

\section{Conclusion}
An automated data labeling and curation workflow has been developed and implemented, featuring the following components:

\begin{itemize}
    \item Daily transfer and QA/QC of data to LBNL's cluster (BDC)
    \item Generation of R/N labels (NMF-based anomaly detection and isotope identification)
    \item Generation of contextual labels (YOLOv10 for per-frame object detection and identification, Norfair for frame-to-frame tracking)
\end{itemize}

As of July 2025, the curated dataset includes 9 months of data from 13 nodes, with gamma-ray data in list/bin-mode formats, environmental data (temperature, pressure, relative humidity, precipitation), selected camera footage, and corresponding labels. Data and labels are organized in monthly collections per node on BDC. The goal is for the data to ultimately be made available to the research community.

The PANDAWN network is a unique system that enables the acquisition of complex multi-modal datasets. When combined with the presented labeling and curation pipeline, it facilitates the generation of high-quality labeled datasets tailored for the development of advanced algorithms, including R/N detection, object detection and tracking, and source-object attribution. 
As data analysis progresses, it is envisioned that the algorithms and ML models used in the analysis will be integrated into the labeling and curation pipeline to expand both the scope and diversity of the generated labels.

Finally, the availability of such datasets enables the exploration of advanced AI methods, such as LMMs and Time Series Foundational Models, to deliver operationally relevant domain awareness through the automated interpretation of spatio-temporal signals and multi-sensor data fusion. 

\bibliographystyle{unsrt}
\bibliography{refs}

\end{document}